%% file: RLBodyguards.tex
\documentclass{article}

\usepackage{microtype}
\usepackage{graphicx}
\usepackage{subfigure}
\usepackage{booktabs} 
\usepackage{amsmath} 
\usepackage{amssymb}  
\usepackage{xcolor}

\usepackage{hyperref}

\usepackage[accepted]{icml2018}

\icmltitlerunning{Multi-Objective Reward for Teams of Robotic Bodyguards}

\begin{document}

\twocolumn[
\icmltitle{Designing a Multi-Objective Reward Function for Creating Teams of Robotic Bodyguards Using Deep Reinforcement Learning}




\begin{icmlauthorlist}
\icmlauthor{Hassam Ullah Sheikh}{to}
\icmlauthor{Ladislau B{\"o}l{\"o}ni}{to}
\end{icmlauthorlist}

\icmlaffiliation{to}{Department of Computer Science, University of Central Florida, Orlando,  Florida, United States}

\icmlcorrespondingauthor{Hassam Ullah Sheikh}{hassam.sheikh@knights.ucf.edu}

\icmlkeywords{Machine Learning, ICML}

\vskip 0.3in
]



\printAffiliationsAndNotice{} 

\begin{abstract}
We are considering a scenario where a team of bodyguard robots provides physical protection to a VIP in a crowded public space. We use deep reinforcement learning to learn the policy to be followed by the robots. As the robot bodyguards need to follow several difficult-to-reconcile goals, we study several primitive and composite reward functions and their impact on the overall behavior of the robotic bodyguards. 
\end{abstract}

\input{Introduction}
\input{RelatedWork}
\input{Background}

\input{ModelingBodyguard}


\input{Experiments}
\input{Conclusions}

\bibliography{ref}
\bibliographystyle{icml2018}
\end{document}

%% file: Introduction.tex
\section{Introduction}
\label{sec:Introduction}


Recent progress in the field of autonomous robots makes it feasible for robots to interact with multiple humans in public spaces. In this paper, we are considering a human VIP moving in a crowded market environment who is protected from physical assault by a team of bodyguard robots. The robots must take into account the position and movement of the VIP, the bystanders and other robots. Previous work in similar problems relied on explicitly programmed behaviors.




Recent research in deep reinforcement learning~\cite{Sergey-2016-JMLR,Silver-2016-Nature} and imitation learning~\cite{Rahmatizadeh-2018-ICRA} applied to the robotics has raised the possibility that learning algorithms might lead to better algorithms than explicit programming. In this paper we explore deep reinforcement learning approaches for our scenario. We aim to simultaneously learn communication and coordination techniques between the agent and the task oriented behavior.


 We aim to develop a general task framework, which can generalize to other types of desired behaviors, beyond bodyguard protection.

In order to achieve these goals, we need to specify: the environment in which the agents will perform, the environment representation in the robot that forms the basis of learning, the reward function that describes the desired behavior, and the reinforcement algorithms deployed. For the bodyguard task, the design of the reward function is especially challenging, because it is task specific, and it needs to reconcile multiple conflicting objectives – the maximum protection, while minimizing interference with the crowd and being unobtrusive. In Section~\ref{sec:RewardFunctions} we discuss several different reward functions that reflect the different aspects of the desired behavior.

We describe several experiments using the Multi-agent Deep Deterministic Policy Gradient (MADDPG) algorithms over several choices of reward functions. We found that communication penalization reward functions captures better the collaborative nature of the scenario, and thus it performs better in experiments.

%% file: RelatedWork.tex
\section{Related Work}
\label{sec:RelatedWork}

The use of robots as a bodyguards is related to several different area of
research and has received significant attention lately. Several different
studies such as~\cite{Klima-2016-NIPS, Hirofumi-2015-ASCC} considered
using robots and multi-agent reinforcement learning for security related
tasks such as patrolling and team coordination by placing checkpoints
to provide protection against imminent adversaries. A multi-robot
patrolling framework was proposed by~\cite{Saad-2012-FLAIRS}
that analyzes the behavior pattern of the soldiers and the robot and
generates a patrolling schedule. The control of robots for providing
maximal protection to a VIP was well studied in \cite{Bhatia-2016-FLAIRS}
where they introduced the concept of threat vector resolution and
quadrant load balancing.

%% file: Background.tex
\section{Background}
\label{sec:Background}

We consider the problem of providing maximal physical protection as
a standard reinforcement learning setup with $N$ agent interacting
with the environment in $E$ discrete steps using real valued continuous
actions $a_{t}$ such that $a_{t}\in\mathbb{R}^{d}$. At each timestep
$t$, the agents receive an observation $x_{t}$, takes the action
$a_{t}$, and receive a scalar reward $r_{t}.$ Generally, the environment
can be partially observable i.e we may need the entire past history
of observations, action pairs to represent the current state such
that $s_{t}$= $\left(x_{1},a_{1},x_{2},a_{2,}\ldots,x_{t}\right)$.
For our problem, we have assumed that the environment is fully observable
so we will represent $s_{t}=x_{t}$

The behavior of each agent is represented by its own policy $\pi$ that takes the state $s_{t}$ as an input and outputs
a probability distribution over all the actions i.e $\pi\left(s\right)\rightarrow\mathbb{P}\left(\mathcal{A}\right)$.
Since the environment is stochastic, we will model it as Markov Decision
Process with a state space $\mathcal{S}$, action space $\mathcal{A}$,   reward function $\mathcal{R}\left(s_{t},a_{t},s_{t+1}\right)$ and
the transition dynamics $p(s_{t+1}|s_{t},a_{t})$.

The return $\mathcal{G}_{t}$ from state $s$ at timestep $t$ is defined as
the discounted cummulative reward that the agent accumulates starting
from state $s$ at timestep $t$ and represented as $\mathcal{G}_{t}={\displaystyle \sum_{i=t}^{T}\gamma^{i-t}}$$\mathcal{R}\left(s_{i},a_{i},s_{i+1}\right)$
where $\gamma$ is the discounting factor $\gamma\in\left[0,1\right]$.
The goal of the reinforcement learning is find an optimum policy $\pi^{*}$
that maximizes the expected return starting from state  $s.$ We denote
the trajectory for state visitation for the policy $\pi$ as $\rho^{\pi}.$

%
%

\subsection{Policy Gradients}

Policy gradient methods have been shown to learn the optimal policy in variety of reinforcement learning tasks. The main idea
behind policy gradient methods is instead of parameterizing the Q-function
to extract the policy, we parameterize the policy using the parameters
$\theta$ to maximize the objective which is represented as $J\left(\theta\right)=\mathbb{E}\left[\mathcal{G}_{t}\right]$
by taking a step in the direction of $\nabla J\left(\theta\right)$
where $\nabla J\left(\theta\right)$ is defined as:
\[
\nabla J\left(\theta\right)=\mathbb{E}\left[\nabla_{\theta}\log\pi_{\theta}\left(a|s\right)Q^{\pi}\left(s,a\right)\right]
\]
The policy gradient methods are prone to high variance problem. Several
different methods such as~\cite{Wu-2018-ICLR, Schulman-2017-CORR}
have been shown to reduce the variability in policy gradient methods
by introducing a critic which is a Q-function that tells about the
goodness of a reward by working as a baseline.

\subsection{Deep Deterministic Policy Gradients}

In~\cite{Silver-2014-ICML} has shown that it is possible to extend policy gradient framework to deterministic policies {\em i.e.} $\pi_{\theta}:S\rightarrow\mathcal{A}$.
In particular we can write $\nabla J\left(\theta\right)$
as
\[
\nabla J\left(\theta\right)=\mathbb{E}\left[\nabla_{\theta}\pi\left(a|s\right)\nabla_{a}Q^{\pi}\left(s,a\right)|_{a=\pi\left(s\right)}\right]
\]
Deep Deterministic Policy Gradients~\cite{Lillicrap-2015-ICLR} is an off-policy algorithm and is a modification of the DPG method introduced in \cite{Silver-2014-ICML} in which the policy $\pi$ and the critic $Q^{\pi}$ is approximated using deep neural networks. DDPG also uses an experience
replay buffer alongside with a target network to stabilize the training.

\subsection{Multi-Agent Deep Deterministic Policy Gradients}

Multi-agent deep deterministic policy gradients~\cite{Lowe-2017-NIPS}
is the extension of the DDPG for the multi-agent setting where each
agent has it's own policy. The gradient $\nabla J\left(\theta_{i}\right)$ of
each policy is written as
\[
\nabla J\left(\theta_{i}\right)=\mathbb{E}\left[\nabla_{\theta_{i}}\log\pi_{i}\left(a_{i}|s_{i}\right)Q_{i}^{\pi}\left(s_{i},a_{1},\ldots,a_{N}\right)\right]
\]
\noindent where $Q_{i}^{\pi}\left(s_{i},a_{1},\ldots,a_{N}\right)$ is a centralized action-value function that takes the actions of all the agents in
addition to the state of the agent to estimate the Q-value for agent
$i$. Since every agent has it's own Q-function, its allows the agents
to have different action space and reward functions. The primary motivation
behind MADDPG is that knowing all the actions of other agents makes
the environment stationary, even though their policy changes.

\subsection{Policy Gradients}

Policy gradient methods have been shown to learn the optimal policy in variety of reinforcement learning tasks. The main idea
behind policy gradient methods is instead of parameterizing the Q-function
to extract the policy, we parameterize the policy using the parameters
$\theta$ to maximize the objective which is represented as $J\left(\theta\right)=\mathbb{E}\left[\mathcal{G}_{t}\right]$
by taking a step in the direction of $\nabla J\left(\theta\right)$
where $\nabla J\left(\theta\right)$ is defined as:
\[
\nabla J\left(\theta\right)=\mathbb{E}\left[\nabla_{\theta}\log\pi_{\theta}\left(a|s\right)Q^{\pi}\left(s,a\right)\right]
\]
The policy gradient methods are prone to high variance problem. Several
different methods such as~\cite{Wu-2018-ICLR, Schulman-2017-CORR}
have been shown to reduce the variability in policy gradient methods
by introducing a critic which is a Q-function that tells about the
goodness of a reward by working as a baseline.

\subsection{Deep Deterministic Policy Gradients}

In~\cite{Silver-2014-ICML} has shown that it is possible to extend policy gradient framework to deterministic policies {\em i.e.} $\pi_{\theta}:S\rightarrow\mathcal{A}$.
In particular we can write $\nabla J\left(\theta\right)$
as
\[
\nabla J\left(\theta\right)=\mathbb{E}\left[\nabla_{\theta}\pi\left(a|s\right)\nabla_{a}Q^{\pi}\left(s,a\right)|_{a=\pi\left(s\right)}\right]
\]
Deep Deterministic Policy Gradients~\cite{Lillicrap-2015-ICLR} is an off-policy algorithm and is a modification of the DPG method introduced in \cite{Silver-2014-ICML} in which the policy $\pi$ and the critic $Q^{\pi}$ is approximated using deep neural networks. DDPG also uses an experience
replay buffer alongside with a target network to stabilize the training.

\subsection{Multi-Agent Deep Deterministic Policy Gradients}

Multi-agent deep deterministic policy gradients~\cite{Lowe-2017-NIPS}
is the extension of the DDPG for the multi-agent setting where each
agent has it's own policy. The gradient $\nabla J\left(\theta_{i}\right)$ of
each policy is written as
\[
\nabla J\left(\theta_{i}\right)=\mathbb{E}\left[\nabla_{\theta_{i}}\log\pi_{i}\left(a_{i}|s_{i}\right)Q_{i}^{\pi}\left(s_{i},a_{1},\ldots,a_{N}\right)\right]
\]
\noindent where $Q_{i}^{\pi}\left(s_{i},a_{1},\ldots,a_{N}\right)$ is a centralized action-value function that takes the actions of all the agents in
addition to the state of the agent to estimate the Q-value for agent
$i$. Since every agent has it's own Q-function, its allows the agents
to have different action space and reward functions. The primary motivation
behind MADDPG is that knowing all the actions of other agents makes
the environment stationary, even though their policy changes.

%% file: ModelingBodyguard.tex


\section{Problem Formulation}

The setting we are considering for providing maximal physical protection
to the VIP in crowded environment is a cooperative Markov game which
becomes a natural extension of the single agent MDP for multi-agent
systems. A multi-agent MDP is defined as state space $\mathcal{S}$
that decribes all the configurations of all the agents, an action
space $\mathcal{A}$ that describes the action space of every agent
$\mathcal{A}_{1},\ldots,\mathcal{A}_{N}$. The transitions are defined
as $\mathcal{T}=\mathcal{S}_{1}\times\mathcal{A}_{1}\times\ldots\times\mathcal{S}_{N}\times\mathcal{A}_{N}$.
For each agent $i$, the reward function is defined as $r_{i}=\mathcal{R}\left(\mathcal{S}_{i}, \mathcal{A}_{i}\right)$.

In this problem, we are assuming that all bodyguards have same state
space and they are following an identical policy. For this problem
we are considering a finite horizon problem where each episode is terminated
after \textbf{T} steps. Since this problem is a cooperative setting,
the goal of all the agents is to find individual policies that increase
the collected payoff.

\subsection{The Environment Model}

For the emergence of interesting behaviors in a multi-agent setting, grounded communication in a physical environment is considered to be a crucial component. For performing the experiments, we used Multi-Agent Particle Environment~\cite{Mordatch-2017-ARXIV} which is a two-dimensional physically simulated environment in a discrete time and continuous space. The environment consists of N agent and M landmarks, both possessing physical attributes such as location, velocity and size etc. Agents can act and move independently with their own policies.

In addition to the ability of performing physical actions in the environment, the agents also have the ability to utter verbal symbols over the communication channel at every timestamp. The utterances are symbolic in nature and does not carry any meaning. At each timestamp, every agent utters a categorical variable that is observed by every other agent and it is up to the agents to infer a meaning of these symbols during the training time. Every utterance carry a small penalty and the agent can decide not to utter at every timestamp. We denote the utterance which is a one-hot vector by \textbf{c.}

The complete observation of the environment is given by $o=\left[x_{1,\ldots N+M},c_{1,\ldots N}\right]\in O$. The state of each agent is the physical state of all the entities in the environment and verbal utterances of all the agents. Formally, the state of each agent is defined as $s_{i}=\left[x_{j,\ldots N+M}, c_{k,\ldots N}\right]$ where $x_{j}$ is the observation of the entity $j$ from the perspective of agent $i$ and $c_{k}$ is the verbal utterance of
the agent $k$.

\input{RewardFunctions}

%% file: RewardFunctions.tex
\subsection{Reward Functions}
\label{sec:RewardFunctions}

In~\cite{Bhatia-2016-FLAIRS} has defined a metric that quantifies the threat to the VIP from each crowd member $b_{i}$ at each timestep $t$. This metric can be extended to a reward function. Since the threat level metric gives a probability. We can conclude that when the distance between the VIP and the crowd member is 0, the threat to the VIP is maximum, i.e 1, conversely when the distance between the VIP and the crowd member $b_{i}$ is more than the \textit{safe distance}, the threat to the VIP is 0. We can model this phenomenon as an exponential decay. Thus the fundamental reward function can be defined as
\begin{equation}
\label{eq:threat_level}
\mathcal{R}_{t} \left(\mathcal{B}, \mathit{VIP} \right)=-1+\prod_{i=1}^{k}\left(1-\mathit{TL}\left(\mathit{VIP},b_{i}\right)\right)
\end{equation}
\noindent where
\begin{equation}
\mathit{TL}\left(\mathit{VIP},b_{i}\right)=\exp^{-A\left(\mathit{Dist}(\mathit{VIP},b_{i})\right)/B}
\end{equation}
In the following we will derive reward functions and explain the motivation behind them that were derived from equation~\ref{eq:threat_level}.

The baseline {\em Threat-Only Reward Function} penalizes each agent with the threat to the VIP at each time step as mentioned in~\cite{Bhatia-2016-FLAIRS}.
\begin{equation}
\mathcal{R}_{t} \left(\mathcal{B}, \mathit{VIP}\right)=-1+\prod_{i=1}^{k}\left(1-\mathit{TL}\left(\mathit{VIP},b_{i}\right)\right)
\label{eq:ThreatOnly}
\end{equation}

The {\em Binary Threat Reward Function} penalizes each agent for the threat with a negative binary reward, in addition, each agent is also penalized for not maintaining a suitable distance from the VIP:
\begin{equation}
\mathcal{L}\left(\mathit{VIP},\mathcal{B}\right)=\protect\begin{cases}
-1 & \text{if\,}  \displaystyle -1+\prod_{i=1}^{k}\left(1-\mathit{TL}\left(\mathit{VIP},b_{i}\right)\right) \neq 0
\\
0 & \text{otherwise}
\end{cases}
\end{equation}
\begin{equation}
\mathcal{D}\left(\mathit{VIP},x_{i}\right)=\begin{cases}
0 & m \leq\left\Vert x_{i}-\mathit{VIP}\right\Vert _{2}\leq d\\
-1 & \text{otherwise}\\
\\
\end{cases}
\end{equation}
\noindent where $m$ is the minimum distance the bodyguard has to maintain from VIP and $d$ is the safe distance. The final reward function is represented as
\begin{equation}
\mathcal{R}_{t} \left(\mathcal{B}, \mathit{VIP}, x_i \right)=\mathcal{L}\left(\mathit{VIP},\mathcal{B}\right)+\mathcal{D}\left(\mathit{VIP},x_{i}\right)
\end{equation}

The {\em Composite Reward Function} is the composition of the threat only reward function and the penalty for not maintaining a suitable distance from the VIP
\begin{equation}
\begin{aligned}
\mathcal{R}_{t} \left(\mathcal{B}, \mathit{VIP}, x_i \right)=&\displaystyle -1+\prod_{i=1}^{k}\left(1-\mathit{TL}\left(\mathit{VIP},b_{i}\right)\right) \\&
+\mathcal{D}\left(\mathit{VIP},x_{i}\right)
\end{aligned}
\label{eq:Composite}
\end{equation}

The {\em Communication Penalization Reward Function} augments the composite reward by adding a small penalty $p$ every time the bodyguard performs an utterance, as recommended in ~\cite{Mordatch-2017-ARXIV}.


%% file: Experiments.tex
\section{Experiments}
\label{sec:Experiments}

We performed our experiments using the Multi-Agent Particle Environment (MPE)~\cite{Mordatch-2017-ARXIV}. The performance was measured using the threat
metric defined in~\cite{Bhatia-2016-FLAIRS} over one episode. The experiments were performed with 2-4 bodyguards ranging from 2-4, and a constant number of 10 bystanders. For all of the experiments, we have trained the agents for 10,000 episodes and limiting the length of the episode to 25 steps.

\begin{figure}[h!]
  \centering
\fbox{\includegraphics[height=0.47\columnwidth]{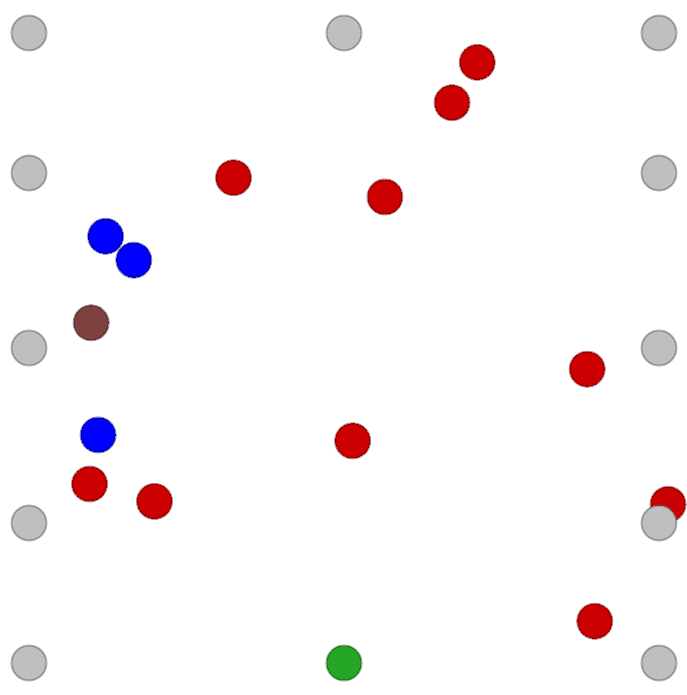}}
\fbox{\includegraphics[height=0.47\columnwidth]{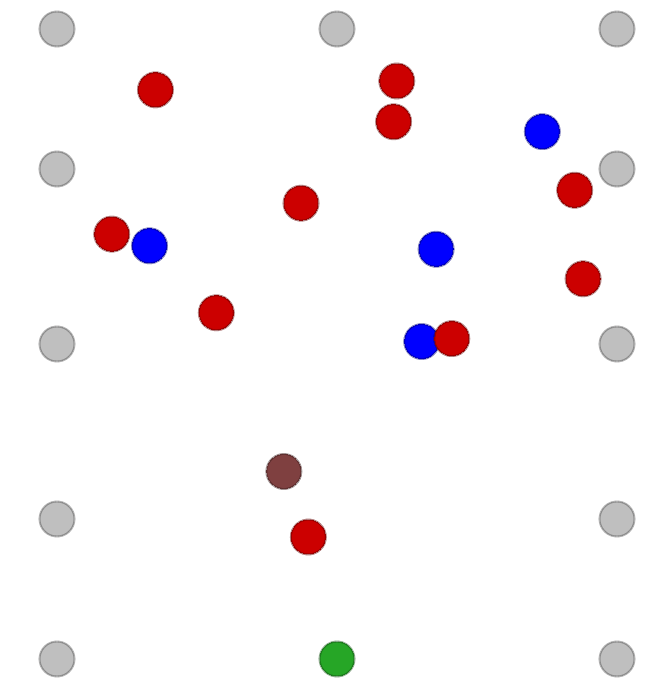}}
\caption{The emerging collaborative bodyguard behavior, using the composite reward function from Equation~\ref{eq:Composite} (left) and the threat-only function from Equation~\ref{eq:ThreatOnly} (right). The VIP is brown, bodyguards blue, bystanders red and landmarks grey.}
\label{fig:screenshots}
\end{figure}

Figure~\ref{fig:screenshots} shows examples of the resulting bodyguard behavior for the composite reward function (left) and the threat only reward function (right).  Notice that for the threat only behavior, the bodyguards are not in the close proximity of the VIP - they have found ways to keep the threat low by ``attacking'' the crowd.



\begin{figure}[h!]
\includegraphics[width=\columnwidth]{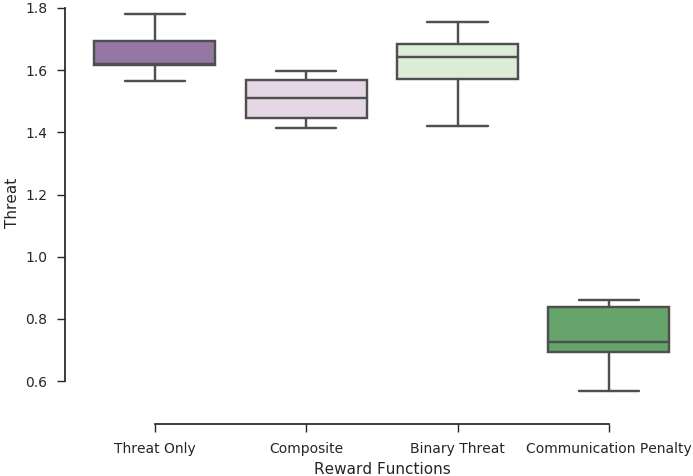}
\vspace{-20pt}
\caption{The overall threat level achieved by agents trained using 4 different reward functions (threat only, composite, binary threat and communication penalty). The scenario involved 4 agents and 10 bystanders.}
\label{fig:reward_function_comparison}
\end{figure}

Figure~\ref{fig:reward_function_comparison} shows the threat levels obtained by different reward functions. The communication penalty function appears the clear winner, with the lowest threat level obtained over the course of the scenario.

%% file: Conclusions.tex


\noindent{\bf Acknowledgement:}  This research was sponsored by the Army
Research Laboratory and was accomplished under Cooperative Agreement Number
W911NF-10-2-0016. The views and conclusions contained in this document are
those of the author’s and should not be interpreted as representing the
official policies, either expressed or implied, of the Army Research
Laboratory or the U.S. Government. The U.S. Government is authorized to
reproduce and distribute reprints for Government purposes notwithstanding
any copyright notation herein.